\documentclass[journal, hidelinks]{IEEEtran}
\usepackage{blindtext}
\usepackage{graphicx}
\usepackage{booktabs}
\usepackage{hyperref}
\usepackage{multirow}
\usepackage{color,url}
\usepackage{enumerate}
\usepackage[ruled,linesnumbered]{algorithm2e}
\usepackage[labelfont=bf]{caption}
\usepackage{subfigure}

\usepackage[cmex10]{amsmath}
\usepackage{array}

\begin{document}
\title{Phone-Level Prosody Modelling with GMM-Based MDN for Diverse and Controllable Speech Synthesis}

\author{Chenpeng Du,~\IEEEmembership{Student Member,~IEEE,}
        and Kai Yu,~\IEEEmembership{Senior Member,~IEEE}
}

\markboth{IEEE/ACM TRANSACTIONS ON AUDIO, SPEECH, AND LANGUAGE PROCESSING}
{Du \MakeLowercase{\textit{et al.}}: Phone-Level Prosody Modelling with GMM-Based MDN for Diverse and Controllable Speech Synthesis}

\maketitle

\begin{abstract}
Generating natural speech with a diverse and smooth prosody pattern is a challenging task. Although random sampling with phone-level prosody distribution has been investigated to generate different prosody patterns, the diversity of the generated speech is still very limited and far from what can be achieved by humans. This is largely due to the use of uni-modal distribution, such as single Gaussian, in the prior works of phone-level prosody modelling. In this work, we propose a novel approach that models phone-level prosodies with a GMM-based mixture density network(MDN) and then extend it for multi-speaker TTS using speaker adaptation transforms of Gaussian means and variances. 
 Furthermore, we show that we can clone the prosodies from a reference speech by sampling prosodies from the Gaussian components that produce the reference prosodies. Our experiments on LJSpeech and LibriTTS dataset show that the proposed method with GMM-based MDN not only achieves significantly better diversity than using a single Gaussian in both single-speaker and multi-speaker TTS, but also provides better naturalness. The prosody cloning experiments demonstrate that the prosody similarity of the proposed method with GMM-based MDN is comparable to recent proposed fine-grained VAE while the target speaker similarity is better.

\end{abstract}


\IEEEpeerreviewmaketitle

\section{Introduction}
\label{sec:intro}

\IEEEPARstart{T}{ext-to-speech} (TTS) synthesis is a process that transforms a transcript into its corresponding speech. Traditional statistical parametric speech synthesis (SPSS) \cite{spss,hmm_dnn_tts} typically contains multiple components, such as a text front-end, a duration model and an acoustic model. In recent years, end-to-end speech synthesis that directly synthesizes speech from text is widely explored based on deep learning. For example, sequence-to-sequence models, such as tacotron2 \cite{tacotron2} and transformer TTS \cite{transformertts}, are able to generate highly natural speech that is comparable to human speech. Besides, non-autoregressive TTS models are also investigated for stability and fast generation, including FastSpeech2 \cite{fastspeech2} and parallel 
Tacotron2 \cite{parallel_tacotron2}.

Besides the linguistic content, speech contains a lot of non-linguistic or para-linguistic information such as speaker identity and prosody. Typically, prosody refers to intonation, speed, intensity, and etc \cite{prosody_define}, which significantly affects the naturalness of speech. Prosody has a top-down hierarchical structure \cite{prosody_hierarchy,prosody_hierarchy2,lhy_hierarchy}, including discourse, phrase, utterance, word, syllable and phone. 
In traditional TTS systems, prosodic features, especially pitch and duration, are exhaustively investigated for generating naturally sounding speech. \cite{tradition_pitch1} proposes a rule-based prosody prediction method while \cite{tradition_pitch2} generates the prosody with a weighted finite-state transducer (WFST). \cite{tradition_pitch3} applies the restriction of natural pitch range to the unit selection synthesizer and improves the intonation naturalness. There are also researches exploiting prosodic features for emotional speech synthesis \cite{tradition_emo1,tradition_emo2}.

In recent end-to-end speech synthesis, literatures mainly focus on utterance-level \cite{ref_speech,gst,tacotron_vae} and phone-level \cite{phone_level_prosody_amazon,phone_level_3dim_quantize,phone_level_3dim} prosody, where the prosody representations are extracted by neural networks. Prosody modelling is important for generating diverse speech corresponding to the same input text. Given the distribution of prosody representations, we can sample various prosodies from the distribution to guide the speech synthesis. Compared with utterance-level prosody modelling, phone-level prosody is more fine-grained to precisely control synthesised speech. Hence, we focus on phone-level prosody modelling in this work.

 Additionally, prosody cloning and prosody transfer are two downstream tasks that controls the prosody with a reference in synthesizing speech of a target speaker. In most cases, the training corpus is in the form of text audio pairs without prosody labelling. Therefore, we often describe the prosody with a reference speech and try to imitate the reference prosody in speech synthesis. There are two different cases: a) the reference speech is constrained to share the same linguistic content with the input text \cite{ref_speech,prosody_transfer,copycat}, b) no constraint is imposed \cite{gst,tacotron_vae}. To avoid confusion, we refer to the two cases as {\em prosody cloning} and {\em prosody transfer} respectively.
In this work, we only focus on the prosody cloning task. One of the recent popular methods for prosody cloning is fine-grained variational auto-encoder (VAE) \cite{prosody_transfer,copycat} which extracts prosody representations from the reference speech with a fine-grained VAE architecture for guiding the synthesis.

Despite the success of the prior works for phone-level prosody modelling, it is still challenging to generate highly diverse speech. This is largely due to the use of uni-modal distribution, such as single Gaussian. Actually, phone-level prosodies are highly diverse even for the same context, hence it is natural to apply multi-modal distribution. In traditional automatic speech recognition (ASR) systems, one of the most dominant techniques is HMM-GMM \cite{gmmasr1,gmmasr2,gmmasr3}, in which the distribution of acoustic features for each HMM state is modeled with a GMM. Similarly, GMM is also used to model acoustic features in traditional SPSS \cite{tts_mdn1,tts_mdn2} and has improved voice quality.

To achieve more accurate and controllable prosody representation, we propose a novel approach that models phone-level prosodies with a GMM-based mixture density network (MDN) \cite{mdn}.
We use a prosody extractor to extract phone-level prosody embeddings from ground-truth mel-spectrograms and use a prosody predictor of MDN to predict the GMM distribution of the embeddings. 
In inference stage, the prosody of each phone is randomly sampled from the predicted GMM distribution for generating speech with diverse prosodies. 
Furthermore, we extend the GMM-based prosody modelling for multi-speaker TTS. For each GMM distribution, the speaker-independent means and variances of all Gaussian components are nonlinearly transformed to speaker-dependent ones with a same group of parameters.
The proposed TTS model with GMM-based phone-level prosody modelling is called PLP-GMM in the rest of this paper, while the model with single Gaussian modelling is called PLP-SG. We also use a baseline that models utterance-level prosodies with a VAE, which is called ULP. Our experiments are performed on single-speaker LJSpeech \cite{ljspeech} and multi-speaker LibriTTS \cite{libritts} dataset respectively. We find that PLP-GMM not only achieves better diversity than PLP-SG and ULP in both single-speaker and multi-speaker tasks, but also provides better naturalness than PLP-SG.

Furthermore, in PLP-GMM, phone-level prosodies are clustered into Gaussian components. In other words, the Gaussian component index represents a certain type of prosody. Hence we can control the prosodies of synthetic speech by sampling phone-level prosody embeddings from the specified components. Therefore, it is natural to clone the reference prosody by synthesizing speech with specific components that produce the reference prosody embeddings. Our experiments show that the prosody similarity of the proposed GMM-based cloning method is comparable to the fine-grained VAE while the target speaker similarity is better.

The main contributions of this work are as follows:
\begin{itemize}
    \item We propose a novel approach that utilizes a GMM-based mixture density network for phone-level prosody modelling in end-to-end speech synthesis, which significantly improves the diversity of generated prosody without losing naturalness.
    \item We propose speaker adaptation with nonlinear transformations of Gaussian means and variances to effectively apply GMM-based phone-level prosody modelling for multi-speaker TTS.
    \item We propose a GMM-based prosody cloning method, which achieves better target speaker similarity than prior works.
\end{itemize}

In the rest of this paper, prior works of prosody modelling and cloning are reviewed in Section \ref{sec:review}. Then we introduce the proposed GMM-based prosody modelling method for single-speaker TTS in Section \ref{sec:pl_prosody_modelling_ss} and for multi-speaker TTS in Section \ref{sec:pl_prosody_modelling_ms}. GMM-based prosody cloning scheme is described in Section \ref{sec:prosody_cloning}. Section \ref{sec:setup} and \ref{sec:exp_result} gives experiments setup and results, and finally Section \ref{sec:conclusion} concludes the paper.

\section{Related work}
\label{sec:review}

\subsection{Prosody modelling}

Utterance level prosody modelling in TTS is first investigated in \cite{ref_speech}, in which a global (utterance-level) prosody embedding is extracted from a reference speech for controlling the prosody of TTS output. 
\cite{gst} factorizes the prosody embedding with several global style tokens(GST), each of which represents a certain speaking style, whose weighted sum specifies the prosody. 
Variational auto-encoder(VAE) is also used for prosody modelling in \cite{tacotron_vae}.
However, all of the above methods can only describe prosodies at utterance-level, while phone-level prosodies can control the generated speech more precisely. Hence, in this work, we only focus on phone-level prosody modelling.

Phone-level prosody modelling is analyzed in several recent works.
\cite{prosody_attention} tries to use an attention module to align the reference frame-level prosodies with each phone. \cite{phone_level_prosody_amazon} finds the instability problem in this method, so it directly extracts phone-level prosody from the segment of acoustic features corresponding to each phone, and models the posterior distribution of the extracted prosody with a single Gaussian in a VAE architecture. In inference stage, the prosodies of phones are independently sampled from the prior distribution $\mathcal{N}(0,\mathbf{I})$. \cite{phone_level_3dim_quantize} auto-regressively predicts the phone-level prosody distribution for each phone, and conditions the prediction on the history of phone-level prosodies with an LSTM. This can be formulated as
\begin{equation}
\begin{aligned}
\label{eq:pl_sg}
p\left(\mathbf{e}_k;\mathbf{e}_{<k},\mathbf{X}\right) = \mathcal{N}\left(\mathbf{e}_k ; \boldsymbol{\mu}_k, \boldsymbol{\sigma}^{2}_k\right)
\end{aligned}
\end{equation}
where $\mathbf{e}_{k}$ represents the prosody of the $k$-th phone and $\mathbf{X}$ represents the input phone sequence.
\cite{phone_level_3dim} proposes hierarchical VAE for phone-level prosody modelling, which tries to obtain an interpretable latent space.

However, most of the prior works for phone-level prosody modelling assumes that the distribution of prosody embeddings is a single Gaussian, which does not have sufficient complexity to model rich prosodies. This leads to limited prosody diversity of generated speech. \cite{du21b_interspeech} proposes a GMM-based prosody modelling approach in single-speaker TTS. In this work, we further propose a GMM-based prosody modelling approach in multi-speaker TTS and apply it to prosody cloning.

\begin{figure}[htbp!]
\centering
\subfigure[Training stage]{
\includegraphics[width=\linewidth,height=2cm]{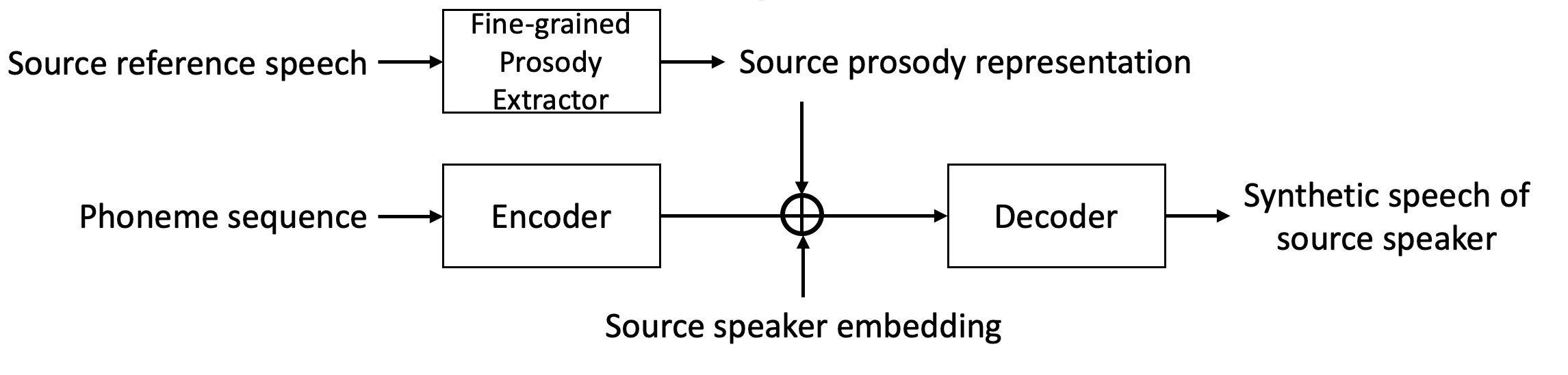}
}
\subfigure[Inference stage]{
\includegraphics[width=\linewidth]{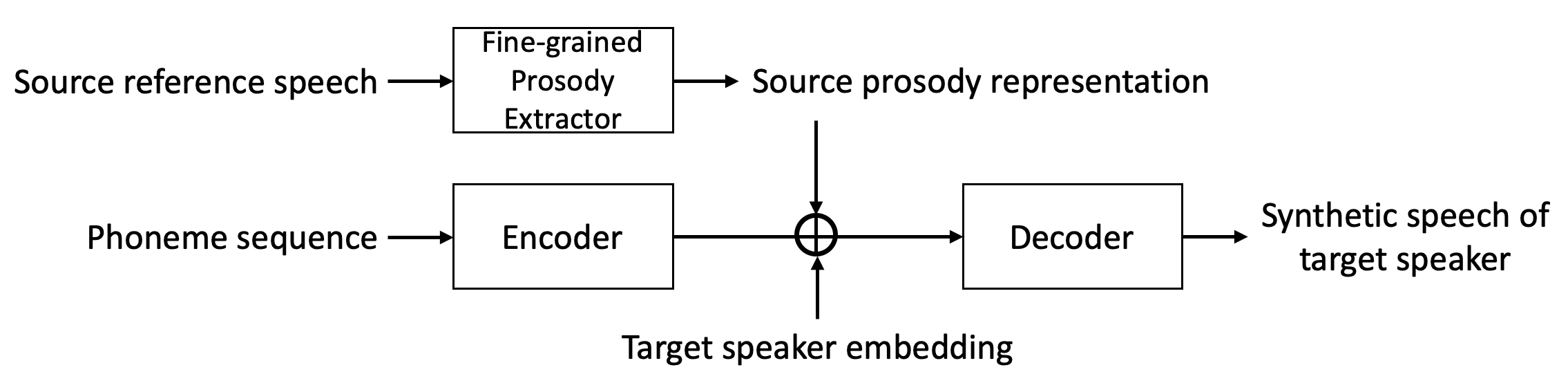}
}
\caption{Prosody cloning pipeline with fine-grained VAE.}
\label{fig:copycat}
\end{figure}

\begin{figure*}[t]
  \centering
  \subfigure[Overal architecture based on FastSpeech2]{
    \includegraphics[height=7.6cm]{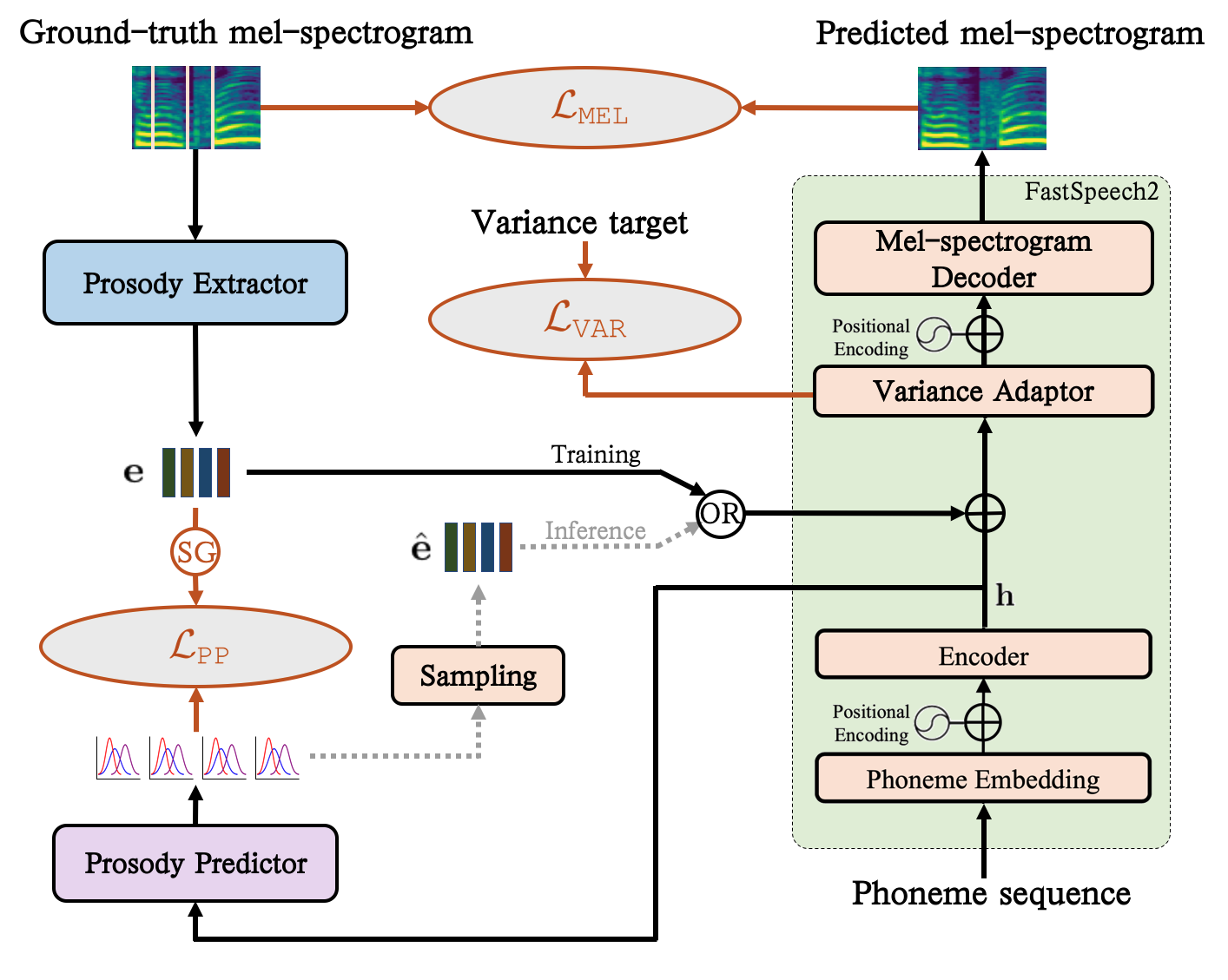}
    \label{overall}
  }
  \subfigure[Prosody extractor]{
    \includegraphics[height=6.5cm]{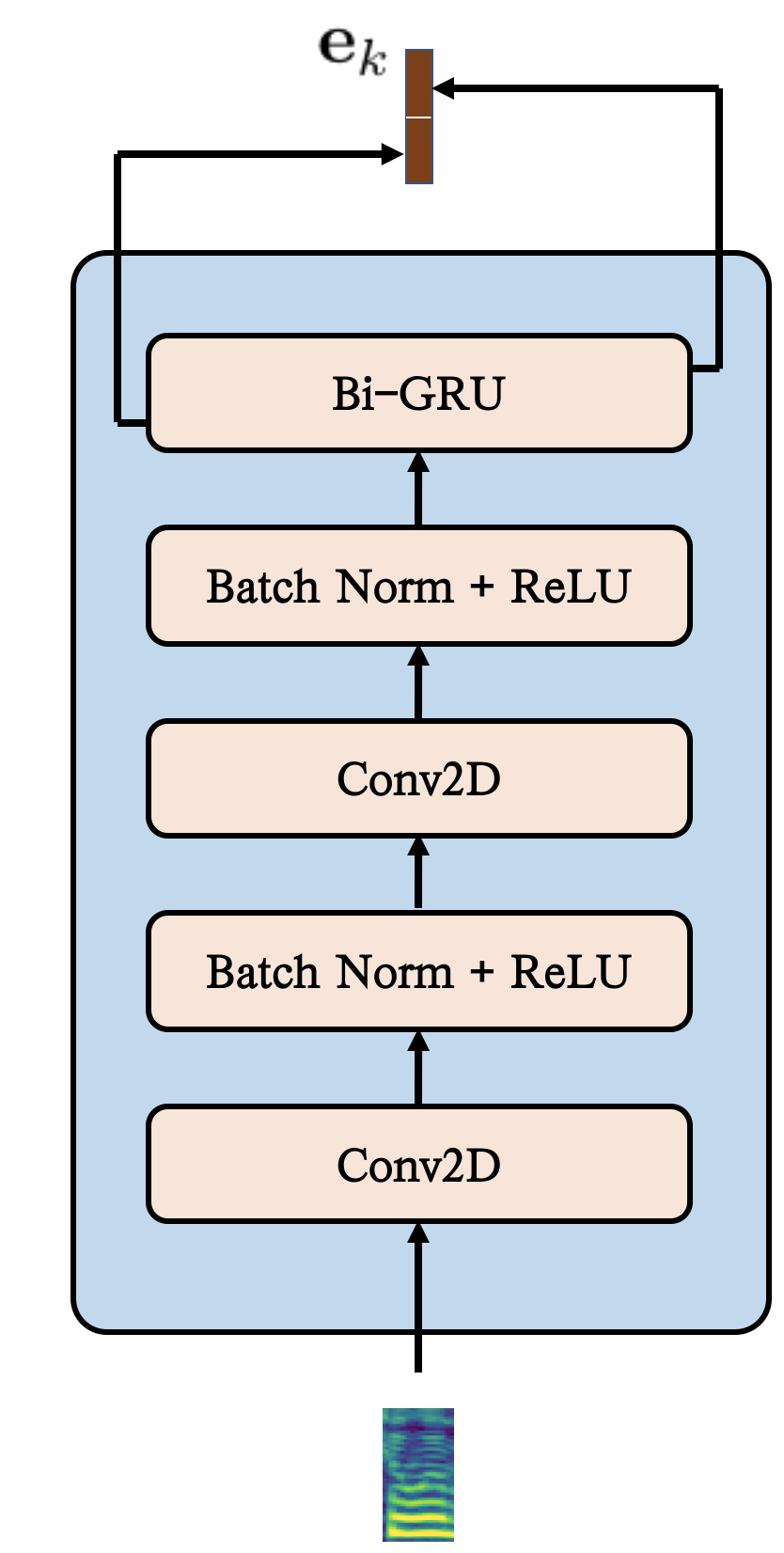}
    \label{extractor}
  }
  \subfigure[Prosody predictor]{
    \includegraphics[height=7.5cm]{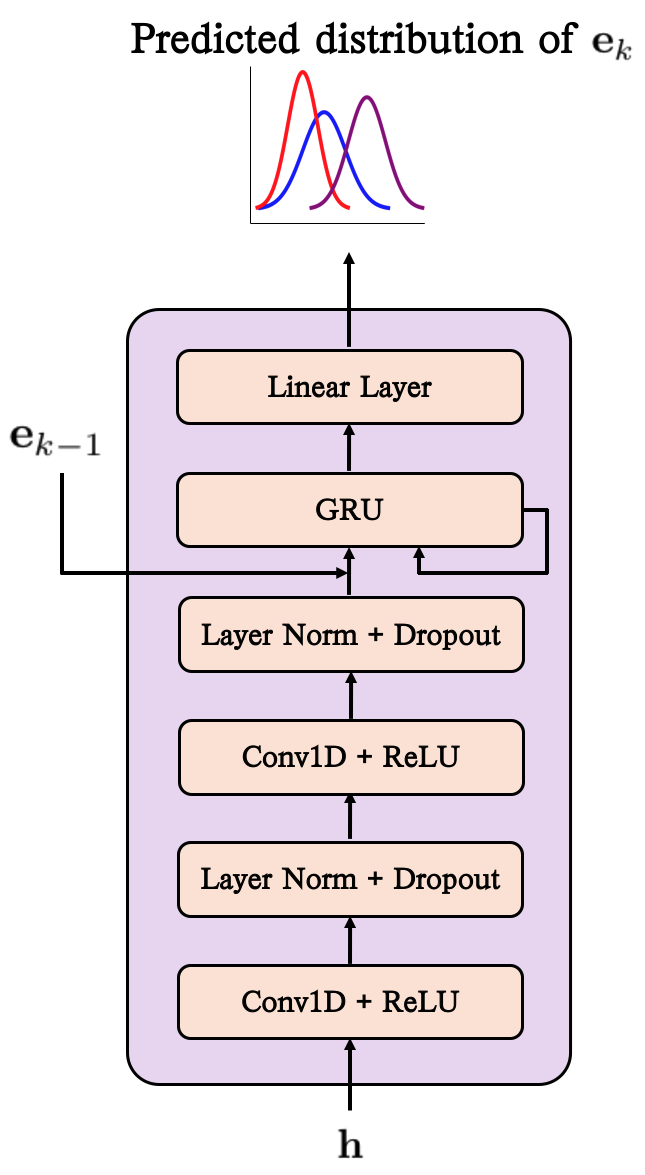}
    \label{predictor}
  }
  \caption{The proposed architectures for single speaker TTS. ``SG'' represents the stop gradient operation. ``OR'' selects the extracted ``ground-truth'' $\mathbf{e}$ in training stage and the sampled $\hat{\mathbf{e}}$ in inference stage. We use red lines for loss calculation and dash lines for sampling.}
  \label{fig:single_spk_model}
\end{figure*}

\subsection{Prosody cloning}
\label{sec:copycat}

Apart from sampling diverse prosodies for speech synthesis, prosody cloning is another important task. It clones the prosody of a source speech to the synthetic speech of a target speaker with a same linguistic content. \cite{ref_speech} extracts an utterance-level prosody embedding from the reference speech for the synthesis. \cite{prosody_transfer,copycat} utilizes a fine-grained VAE for prosody cloning, whose pipeline is shown in Figure \ref{fig:copycat}. In the training stage, the reference speech is exactly the same as the training target. The posterior of latent variables are calculated by a prosody extractor. Then the prosody representations are obtained from the posterior and then concatenated to the encoder output together with the speaker embedding. In the inference stage, the synthesis is conditioned on the target speaker embedding and the prosody representations of the source speech.

However, the prosody representations are prone to speaker identity interference. 
Although \cite{copycat} introduces a normalization convolutional layer and a bottleneck layer to alleviate the problem, the model structure and training strategy still need to be carefully designed and the problem still occurs in some situations. 

In this work, we model the prosodies with GMMs and then clone the prosodies from a reference speech by sampling prosodies from the Gaussian components that produce the reference prosodies, which fundamentally avoids the speaker identity interference.

\section{GMM-based Phone-level Prosody Modelling for Single-Speaker TTS}
\label{sec:pl_prosody_modelling_ss}

\subsection{Mixture Density Network}
\label{sec:mdn}
 Mixture density network (MDN) is defined as the combined structure of a neural network and a mixture-of-expert model. We focus on GMM-based MDN in this work to predict the parameters of the GMM distribution, including means $\boldsymbol{\mu}_{i}$, variances $\boldsymbol{\sigma}_{i}^{2}$, and mixture weights $w_i$. It should be noted that the sum of the mixture weights is constrained to 1, 
which can be achieved by applying a Softmax function, formalized as
\begin{equation}
\begin{aligned}
\label{eq:softmax}
w_{i}=\frac{\exp \left(\alpha_{i}\right)}{\sum_{j=1}^{M} \exp \left(\alpha_{j}\right)}
\end{aligned}
\end{equation}
where $M$ is the number of Gaussian components and $\alpha_{i}$ is the corresponding neural network output. The mean and variance of Gaussian components are presented as
\begin{eqnarray}
\boldsymbol{\mu}_{i}& = & \mathbf{m}_{i} \label{eq:mean}\\
\boldsymbol{\sigma}_{i}^{2}& =& \exp \left(\mathbf{v}_{i}\right)\label{eq:var}
\end{eqnarray}
where $\mathbf{m}_{i}$ and $\mathbf{v}_{i}$ are the neural network outputs corresponding to the mean and variance of the $i$-th Gaussian component. Equation (\ref{eq:var}) constrains the $\boldsymbol{\sigma}_{i}^{2}$ to be positive.

The criterion for training MDN is the negative log-likelihood of the observation $\mathbf{y}$ given its input $\mathbf{x}$. Here we can formulate the loss function as
\begin{equation}
\begin{aligned}
\label{eq:mdn_loss}
\mathcal{L}_{\texttt{MDN}} & = -\log p\left(\mathbf{y};\mathbf{x}\right) \\
& = -\log \left(\sum_{i=1}^{M} w_{i} \ \mathcal{N}\left(\mathbf{y} ; \boldsymbol{\mu}_{i}, \boldsymbol{\sigma}_{i}^{2}\right)\right)
\end{aligned}
\end{equation}
Therefore, given the input $\mathbf{x}$, the mixture density network is optimized to predict GMM parameters $w_i$, $\boldsymbol{\mu}_i$ and $\boldsymbol{\sigma}_i$ that maximize the likelihood of $\mathbf{y}$.

\subsection{Overall architecture}

The overall architecture of the proposed system PLP-GMM is shown in Figure \ref{overall}.
The TTS model in this paper is based on the recent proposed FastSpeech2\cite{fastspeech2}, where the input phone sequence is first converted into a hidden state sequence $\mathbf{h}$ by the encoder and then sent to a variance adaptor and a decoder for predicting the output mel-spectrogram. Compared with the original FastSpeech \cite{fastspeech}, FastSpeech2 is optimized to minimize the mean square error(MSE) $\mathcal{L}_\texttt{MEL}$ between the predicted and the ground-truth mel-spectrograms, instead of applying a teacher-student training. Moreover, the duration target is not extracted from the attention map of an autoregressive teacher model, but from the forced alignment of speech and text. Moreover, \cite{fastspeech2} condition the prediction of mel-spectrogram on the variance information such as pitch and energy with a variance adaptor. The adaptor is trained to predict the variance information with an MSE loss $\mathcal{L}_\texttt{VAR}$.

In this work, we introduce a prosody extractor and a prosody predictor in the FastSpeech2-based TTS system.
In the training stage, the prosody extractors extract a sequence of hidden vectors $\mathbf{e}$ for each input phone from its corresponding mel-spectrogram segment, representing its prosody implicitly. Therefore, we call the hidden vectors prosody embeddings in this work. Let $K$ be the length of the input phone sequence. The output of prosody extractor is a sequence of hidden vector with a same length
\begin{equation}
\begin{aligned}
\label{eq:vec_seq}
\mathbf{e}=[\mathbf{e}_1, \mathbf{e}_2, ..., \mathbf{e}_K]
\end{aligned}
\end{equation}
They are then projected and added to the corresponding hidden state sequence $\mathbf{h}$. We do not need specific training labels for the prosody extractor, for it is jointly trained with the whole TTS model and is optimized to generate the prosody embeddings that can better reconstruct the output mel-spectrogram, which is similar to an auto-encoder.

The extracted prosody embeddings $\mathbf{e}_k$ are used as the target to train the prosody predictor during training. 
The distribution of $\mathbf{e}_k$ is assumed to be a GMM whose parameters are predicted by an MDN. Here, the MDN is the prosody predictor. Let $M$ be the number of Gaussian components.
The prosody predictor autoregressively predicts $M$ means, variances and weights for each phone. In inference stage, we sample the prosodies $\hat{\mathbf{e}}$ from the predicted distributions. 

It should be noted that the prosody embeddings in this work are phone-dependent. We do not apply an adversarial loss to remove the potential phone information in the embeddings. However, it does no harm to our speech synthesis, because the distributions of prosody embeddings are predicted from the input phonemes in the inference stage. The prosody embeddings sampled from a distribution represent different prosodies corresponding to the corresponding phone.

\subsection{Prosody extractor and prosody predictor}

The detailed architecture of the prosody extractor is shown in Figure \ref{extractor}. It contains 2 layers of 2D convolution, each followed by a batch normalization layer and a ReLU activation function. A bidirectional GRU is designed after the above modules. The concatenated forward and backward states from the GRU layer is the output of the prosody extractor, which is referred to as the prosody embedding of the phone.  

Figure \ref{predictor} demonstrates the detailed architecture of the prosody predictor.
The hidden state $\mathbf{h}$ of the input phone sequence is sent to 2 layers of 1D convolution, each followed by a ReLU, layer normalization and dropout layer. The output of the above modules is then concatenated with the previous prosody embedding and sent to a GRU. The GRU is designed to condition the prediction of the current prosody distribution on the previous prosodies. Then we project the GRU output to obtain $w_{k,i}$, $\mathbf{m}_{k,i}$ and $\mathbf{v}_{k,i}$, which is then transformed to the GMM parameters according to Equation (\ref{eq:softmax}) - (\ref{eq:var}).

Equation (\ref{eq:mdn_loss}) formulates the training criterion for an MDN, which is the negative log-likelihood of the observations. Here, the observations are the prosody embeddings $\mathbf{e}$, so we obtain the loss function for training the prosody predictor
\begin{equation}
\begin{aligned}
\label{eq:pp_loss}
\mathcal{L}_{\texttt{PP}} & = \sum_{k=1}^{K} -\log p\left(\mathbf{e}_k;\mathbf{e}_{<k},\mathbf{h}\right) \\
& = \sum_{k=1}^{K} -\log \left(\sum_{i=1}^{M} w_{k,i} \  \mathcal{N}\left(\mathbf{e}_k ; \boldsymbol{\mu}_{k,i}, \boldsymbol{\sigma}_{k,i}^{2}\right)\right)
\end{aligned}
\end{equation} 
where $w_{k,i}$, $\boldsymbol{\mu}_{k,i}$ and $\boldsymbol{\sigma}_{k,i}$ are the GMM parameters of the $i^{th}$ component of phone $k$. They are predicted given $\mathbf{h}$ and $\mathbf{e}_{<k}$.

\begin{figure*}[t]
  \centering
  \subfigure[Overal architecture based on FastSpeech2]{
    \includegraphics[height=7.6cm]{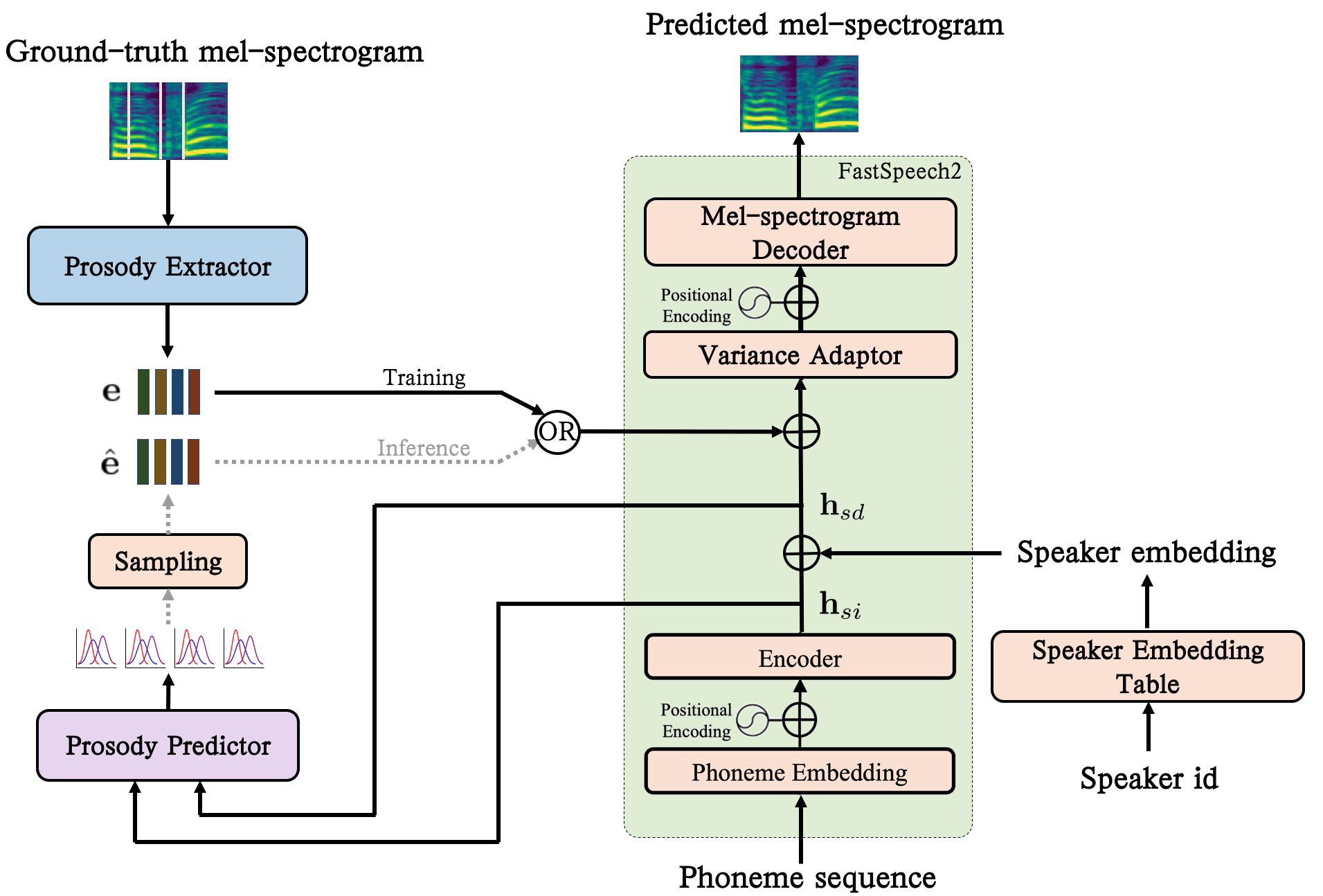}
    \label{ms_overall}
  }
  \subfigure[Prosody predictor]{
    \includegraphics[height=9.5cm]{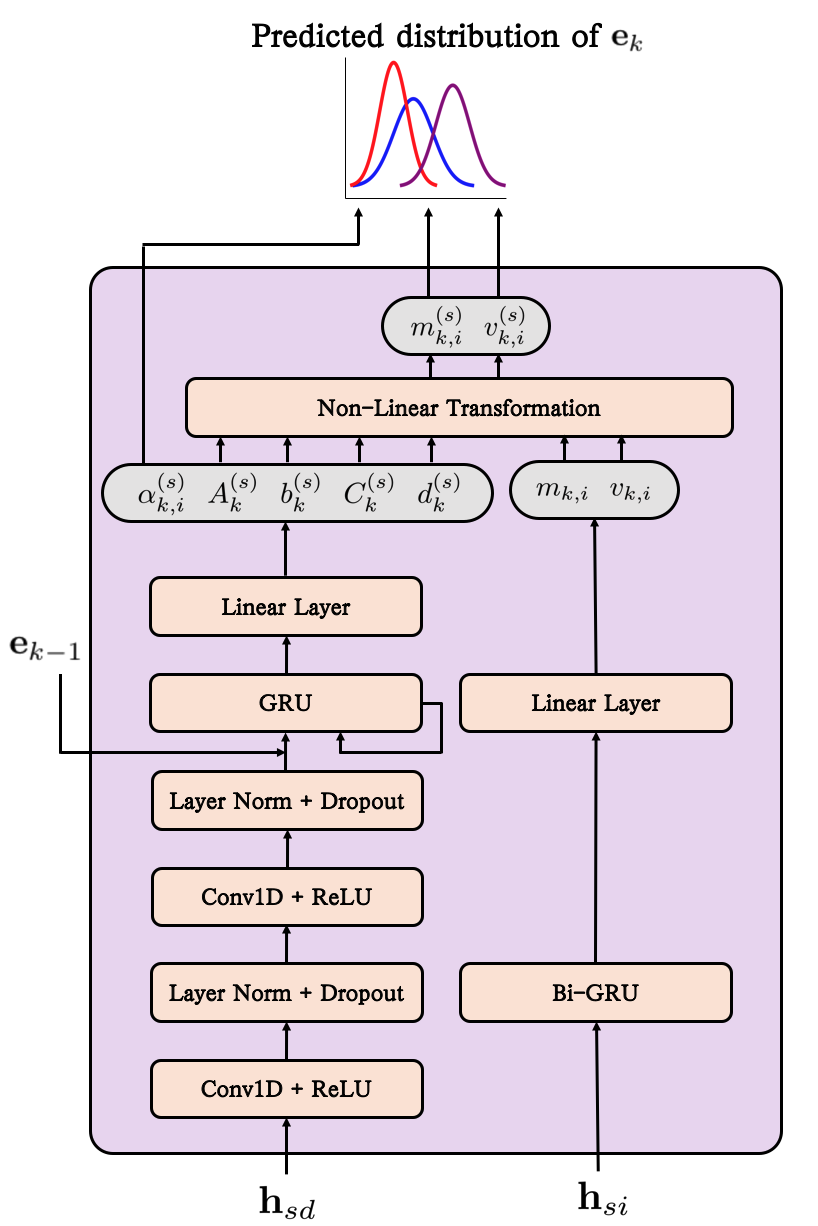}
    \label{ms_predictor}
  }
  \caption{The extension of the proposed architecture for multi-speaker TTS. The loss function for the TTS system is the same as in Figure \ref{overall}, so it is eliminated in \ref{ms_overall} for simplicity.}
  \label{fig:multi_speaker_model}
\end{figure*}

\subsection{Training criterion}

The prosody extractor and the prosody predictor are both jointly trained with the FastSpeech2 architecture. The overall architecture is optimized with the loss function
\begin{equation}
\begin{aligned}
\label{eq:overall_loss}
\mathcal{L} & =  \beta \  \mathcal{L}_{\texttt{PP}} + \mathcal{L}_{\texttt{FastSpeech2}} \\
& = \beta \ \mathcal{L}_{\texttt{PP}} + \left( \mathcal{L}_{\texttt{MEL}} + \mathcal{L}_{\texttt{VAR}} \right)
\end{aligned}
\end{equation}
where $\mathcal{L}_{\texttt{PP}}$ is defined in Equation (\ref{eq:pp_loss}), $\mathcal{L}_{\texttt{FastSpeech2}}$ is the loss function of FastSpeech2 which is the sum of variance prediction loss $\mathcal{L}_{\texttt{VAR}}$ and mel-spectrogram reconstruction loss $\mathcal{L}_{\texttt{MEL}}$ as described in \cite{fastspeech2}, and $\beta$ is the relative weight between the two terms.
It should be noted that we use a stop gradient operation on $\mathbf{e}$ in calculating the $\mathcal{L}_{\texttt{PP}}$, so the prosody extractor is not optimized with $\mathcal{L}_{\texttt{PP}}$ directly. 

\section{Speaker Adaptation of GMM for Multi-Speaker TTS}
\label{sec:pl_prosody_modelling_ms}


In Section \ref{sec:pl_prosody_modelling_ss}, we propose PLP-GMM for single-speaker TTS, in which phone-level prosodies are clustered into $M$ Gaussian components, each of them represents a type of prosody.
In multi-speaker TTS, the speaker embedding is selected from a look up table and added to the speaker-independent(SI) encoder output $\mathbf{h}_{si}$, yielding the speaker-dependent(SD) hidden sequence $\mathbf{h}_{sd}$ for speaker-dependent prosody prediction. Further, we need to maintain the cluster results across speakers in order to control the prosodies with Gaussian indices across speakers.

There is a trivial method that directly predicts the SD means $\mathbf{m}_{k,i}^{(s)}$ and log-variances $\mathbf{v}_{k,i}^{(s)}$ with $\mathbf{h}_{sd}$. However, in this case, there is no constraint that ensures same types of prosodies are clustered in same Gaussian components across speakers. Actually, we do find in our experiments that this strategy often leads to the unstable result across speakers, especially across genders.

Therefore, in this work, we propose a novel method that first predicts the SI means $\mathbf{m}_{k,i}$ and log-variances $\mathbf{v}_{k,i}$ and then non-linearly transforms them to the corresponding SD ones $\mathbf{m}_{k,i}^{(s)}$ and $\mathbf{v}_{k,i}^{(s)}$. Due to the speaker independence of SI parameters, only the prosody is clustered first, which determines the prosody type of each component. The SD parameters transformed from the SI ones contain additional speaker information but inherit the clustering results. Hence, in this case, each Gaussian component represents a same type of prosody across different speakers. The transformation method is detailed below.


We extend the single-speaker TTS in Section \ref{sec:pl_prosody_modelling_ss} to multi-speaker TTS, as is shown in Figure \ref{ms_overall}. Both $\mathbf{h}_{si}$ and $\mathbf{h}_{sd}$ are sent to the prosody predictor.

The prosody predictor for multi-speaker TTS is demonstrated in Figure \ref{ms_predictor}. The SI output, including SI means $\mathbf{m}_{k,i}$ and log-variances $\mathbf{v}_{k,i}$, are obtained from ${\mathbf{h}_{si}}$, while the SD output, including transformation parameters $\mathbf{A}_k^{(s)}$,  $\mathbf{b}_k^{(s)}$,  $\mathbf{C}_k^{(s)}$,  $\mathbf{d}_k^{(s)}$ and the logits of Gaussian components $\alpha_{k,i}^{(s)}$, are obtained from $\mathbf{h}_{sd}$. The architecure for predicting the SD output is the same as in the single speaker system, and an additional bi-directional GRU and a linear projection layer is added for predicting the SI output. In order to calculate SD means $\mathbf{m}_{k,i}^{(s)}$ and log-variances $\mathbf{v}_{k,i}^{(s)}$, we apply a non-linear speaker-dependent transformation to the SI $\mathbf{m}_{k,i}$ and $\mathbf{v}_{k,i}$, which can be formulated as
\begin{eqnarray}
\mathbf{m}_{k,i}^{(s)} = Linear\left(tanh(\ \mathbf{A}_k^{(s)} \mathbf{m}_{k,i}+\mathbf{b}_k^{(s)}\ )\right) \label{eq:mean_transform}\\
\mathbf{v}_{k,i}^{(s)} = Linear\left(tanh(\ \mathbf{C}_k^{(s)} \mathbf{v}_{k,i}+\mathbf{d}_k^{(s)}\ )\right) \label{eq:var_transform}
\end{eqnarray}
It should be noted that all the $M$ Gaussian components are transformed with a same group of parameters $\mathbf{A}_k^{(s)}$,  $\mathbf{b}_k^{(s)}$ and $\mathbf{C}_k^{(s)}$,  $\mathbf{d}_k^{(s)}$. For simplicity, we restrict the $\mathbf{A}_k^{(s)}$ and  $\mathbf{C}_k^{(s)}$ to be diagonal.

\begin{figure*}[t]
\centering
\includegraphics[width=\linewidth]{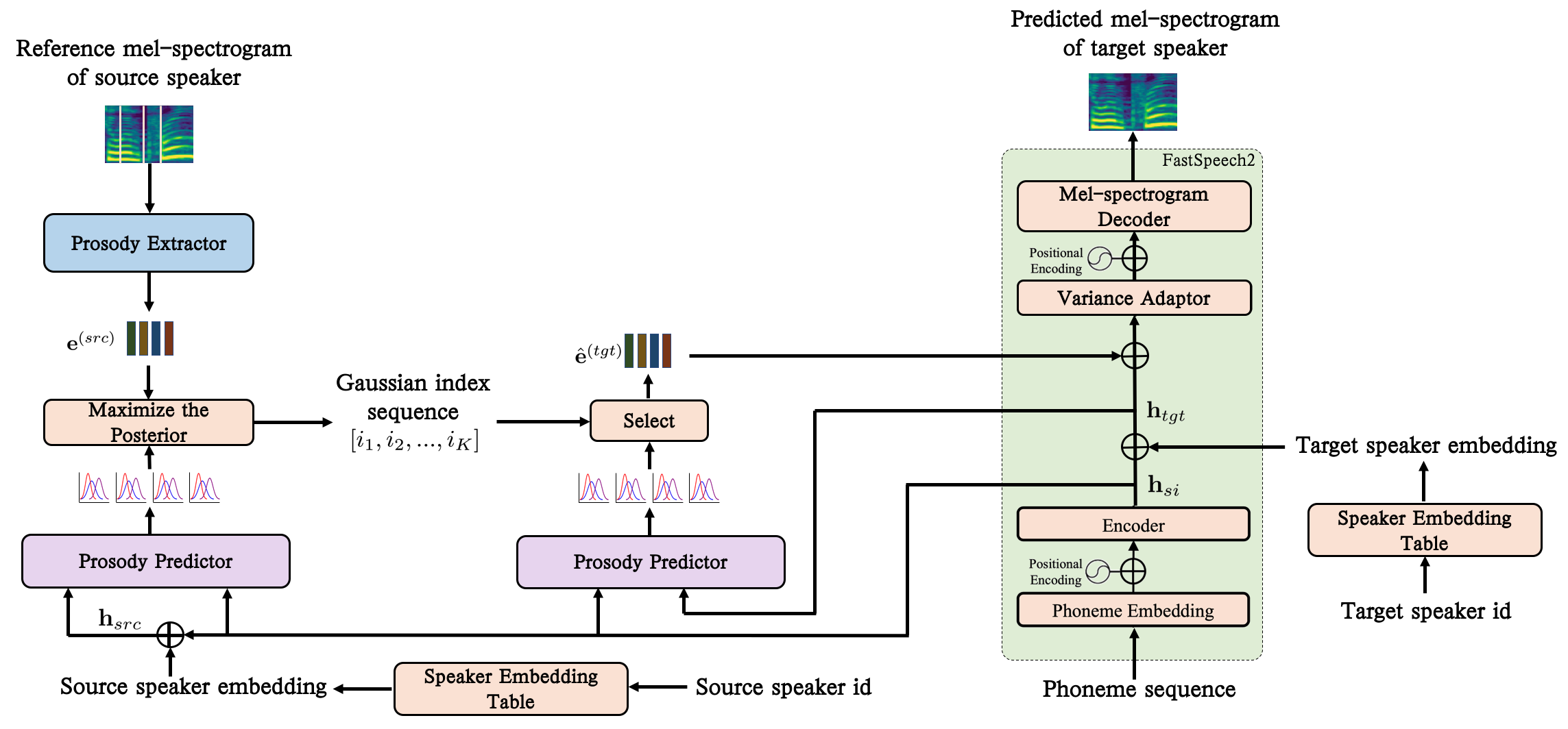}
\caption{Speech Synthesis with cloned phone-level prosodies from a reference speech.}
\label{fig:prosody_cloning}
\end{figure*}

\section{GMM-based Prosody Cloning}
\label{sec:prosody_cloning}

In Section \ref{sec:pl_prosody_modelling_ms}, we propose a GMM-based prosody modelling method for multi-speaker TTS, in which phone-level prosodies are clustered into $M$ Gaussian components. Specifically, the speaker-dependent distributions of different speakers are obtained from a common initial speaker-independent distribution with mean and variance transformations. Here, the transformations are shared across all Gaussian components, so the Gaussian indices represent same clusters after the transformation. Accordingly, although different transformations are applied to different speakers, the Gaussian indices represent same clusters of prosodies across speakers. In other words, the Gaussian component index represents a certain type of prosody, which means we can control the prosodies of synthetic speech by sampling phone-level prosody embeddings from the specified components. 
Therefore, it is natural to clone the reference prosody by synthesizing speech with specific components that produces the reference prosody embeddings.

The pipeline of prosody cloning is demonstrated in Figure \ref{fig:prosody_cloning}. First, we train a multi-speaker TTS system with GMM prosody modelling as described in Section \ref{sec:pl_prosody_modelling_ms}. Then we extract source prosody embeddings $\mathbf{e}^{(src)}$ from the reference speech, predict the Gaussian mixture parameters of the source speaker with the prosody predictor, and then calculate the posterior probability that $\mathbf{e}_k^{(src)}$ comes from the $j$-th Gaussian component. According to the Bayes' Theorem, the posterior probability is
\begin{equation}
\begin{aligned}
\label{eq:bayes}
P\left(j|\mathbf{e}_k^{(src)}\right)=\frac{P\left(j\right)  \  p\left(\mathbf{e}_k^{(src)}|j\right)}{p(\mathbf{e}_k^{(src)})}
\end{aligned}
\end{equation}
The Gaussian component that produces $\mathbf{e}_k^{(src)}$ is the one that maximizes the posterior probability, that is
\begin{equation}
\begin{aligned}
\label{eq:posterior}
i_k & = \arg \!\max_{j} P\left(j|\mathbf{e}_k^{(src)}\right) \\
& = \arg \!\max_{j}\  w_{k,j}^{(src)} \  \mathcal{N}\left(\mathbf{e}_k^{(src)} ; \boldsymbol{\mu}_{k,j}^{(src)}, \boldsymbol{\sigma}_{k,j}^{(src)2}\right)
\end{aligned}
\end{equation}
When $k$ indexes over all $K$ phones, the indices of the Gaussian components that produce $\mathbf{e}^{(src)}$ generate a sequence
\begin{equation}
\begin{aligned}
\label{eq:idx_seq}
\mathbf{i}=[i_1, i_2, ..., i_K]
\end{aligned}
\end{equation}
according to Equation (\ref{eq:posterior}). Then we predict the Gaussian mixture parameters of the target speaker and sample the prosody embedding from the $i_k$-th Gaussian component for the $k$-th phone in speech synthesis, which can be denoted as
\begin{equation}
\begin{aligned}
\label{eq:tgt_prosody}
\hat{\mathbf{e}}_k^{(tgt)} \sim \mathcal{N}\left(\boldsymbol{\mu}_{k,i_k}^{(tgt)}, \boldsymbol{\sigma}_{k,i_k}^{\left(tgt\right)2}\right)
\end{aligned}
\end{equation}
For stability and simplicity, we directly select the mean as the prosody embedding of target speaker, which is
\begin{equation}
\begin{aligned}
\label{eq:tgt_prosody_simple}
\hat{\mathbf{e}}_k^{(tgt)} = \boldsymbol{\mu}_{k,i_k}^{(tgt)}
\end{aligned}
\end{equation}
Thus, the phone-level prosody embeddings $\hat{\mathbf{e}}^{(tgt)}$ from the specified Gaussian indices $\mathbf{i}$ precisely clone the reference prosodies. Further, the GMM parameters for the target speaker are purely predicted with the target speaker embedding, so source speaker interference cannot occur in our prosody cloning.

\section{Experimental Setup}
\label{sec:setup}

\subsection{Dataset}

\subsubsection{Single-speaker dataset}
LJSpeech is an English dataset, containing about 24 hours speech recorded by a female speaker. We randomly leave out 250 utterances for testing.

\subsubsection{Multi-speaker dataset}
 LibriTTS is a multi-speaker English dataset, which consists of 3 parts -- ``train-clean-100", ``train-clean-360" and ``train-other-500". We only use the combination of the two clean parts ``train-clean-100" and ``train-clean-360" in this work, which is called ``train-clean-460''. It contains about 245 hours speech and 1151 speakers. We randomly leave out 378 utterances for testing.

\subsection{Data preparation}
\label{sec:conf_data_prep}
 All the speech data in this work is resampled to 16kHz for simplicity. The mel-spectrograms are extracted with 50ms window, 12.5ms frame shift, 1024 FFT points and 320 mel-bins. Before training TTS, we compute the phone alignment of the training data with an HMM-GMM ASR model trained on Librispeech \cite{librispeech}, and then extract the duration of each phone from the alignment for TTS training.

\begin{figure*}[t]
\centering
\subfigure[LJSpeech training set]{
\includegraphics[width=0.35\linewidth]{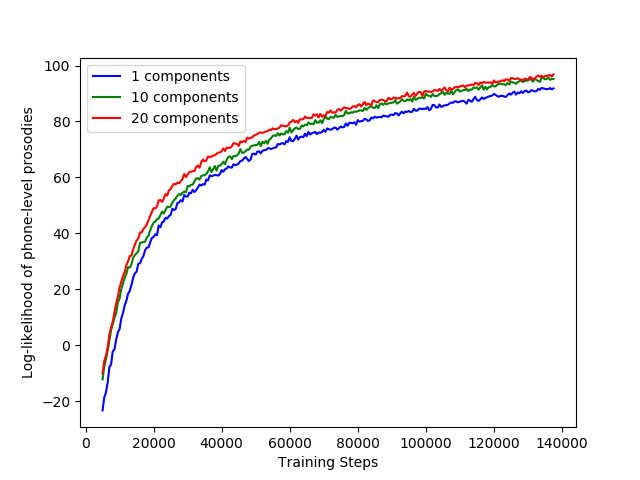}
\label{fig:loglike_ljspeech_train}
}
\subfigure[LJSpeech test set]{
\includegraphics[width=0.35\linewidth]{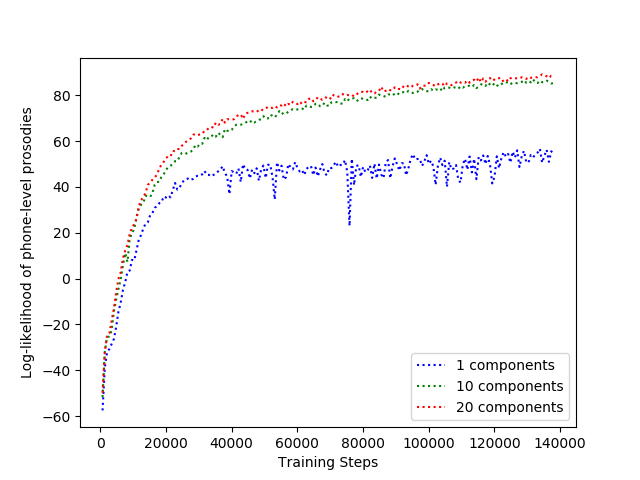}
\label{fig:loglike_ljspeech_valid}
}
\subfigure[LibriTTS training set]{
\includegraphics[width=0.35\linewidth]{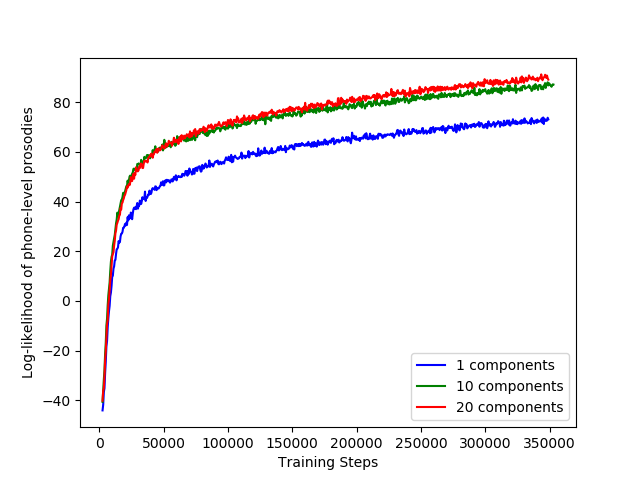}
\label{fig:loglike_libritts_train}
}
\subfigure[LibriTTS test set]{
\includegraphics[width=0.35\linewidth]{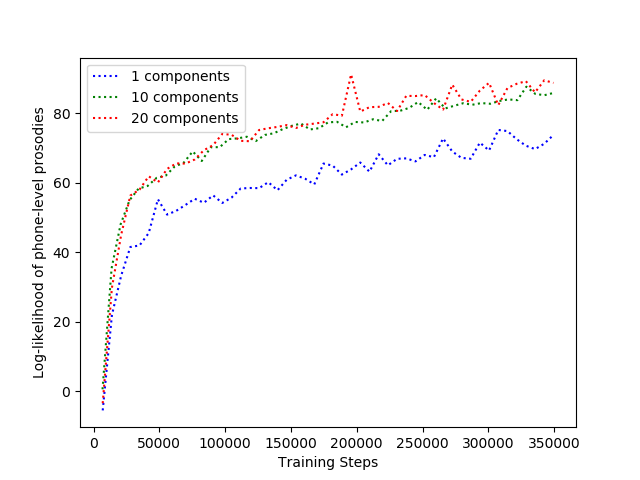}
\label{fig:loglike_libritts_valid}
}
\caption{Log-likelihood curves of the extracted phone-level prosodies with different numbers of Gaussian components on LJSpeech and LibriTTS.}
\label{fig:pitch_contour}
\end{figure*}

\subsection{Acoustic model}
\label{sec:conf_acoustic}
The acoustic model in this work is based on FastSpeech2 \cite{fastspeech2}. It consists of a phone embedding layer, an encoder, a variance adaptor and a decoder.
The phone embedding layer contains a lookup table which is trained together with the TTS system. It transforms the input one-hot phone sequence to a 512 dimensional phone embedding sequence. The encoder consists of 6 layers of Transformer whose output is a 512 dimensional hidden sequence. Two variance adaptors are applied in our system for pitch and energy respectively. They both uses 2 layers of 1D CNN with a kernel size of 3 and a linear projection layer that outputs the predicted pitch and energy. The decoder contains 6 layers of Transformer and a linear projection layer that outputs the mel-spectrogram.
In the multi-speaker TTS, we add a 128 dimensional speaker embedding to the output of the encoder, controlling the speaker identity of the synthetic speech. 

 The $\beta$ in Equation (\ref{eq:overall_loss}) is set to 0.02. An Adam optimizer \cite{adam} is used for TTS training in conjunction with a noam learning rate scheduler \cite{attalluneed}. 
The output mel-spectrogram is converted to the waveform with a MelGAN \cite{melgan} vocoder, which is trained on the same training set.

\subsection{Prosody modelling}
In this section, we describe the configurations of the proposed GMM-based phone-level prosody modelling PLP-GMM and two other prosody modelling methods ULP and PLP-SG. We build three TTS systems with the three prosody modelling methods respectively and they all share the same acoustic model configurations described in Section \ref{sec:conf_acoustic}.

\subsubsection{ULP}
A popular approach for utterance-level prosody (ULP) modelling is conditioning the synthesis on a latent variable from VAE \cite{tacotron_vae}. The dimensionality of the latent variable is set to 128. In the training stage, the latent variable is sampled from the posterior for each utterance. In the inference stage, the latent variable is sampled from the prior distribution, which is a standard Gaussian $\mathcal{N}(0, I)$. 

\subsubsection{PLP-SG}
For precise prosody modelling, we apply phone-level prosodies (PLP) in this work. One basic method of PLP modelling is to use a single Gaussian \cite{phone_level_prosody_amazon,phone_level_3dim_quantize}. 
As shown in Figure \ref{fig:single_spk_model}, we introduce a prosody extractor and a prosody predictor to extract and predict prosody representations for the phones.
 In the prosody extractor, the 2 CNN layers with 8 channels and 3$\times$3 kernel size is followed by a 64 dimensional bi-directional GRU layer. 
 The prosody predictor contains 2 layers of CNN with kernel size 3 and a 512 dimensional GRU followed by a linear projection layer that outputs the parameters of a single Gaussian.
 In multi-speaker TTS, as demonstrated in Figure \ref{fig:multi_speaker_model}, the prosody predictor passes the speaker independent hidden sequence $\mathbf{h}_{si}$ through a 32 dimensional bi-directional GRU followed by a linear projection layer that outputs speaker independent means and log-variances.
 
\subsubsection{PLP-GMM}
As is described in Section \ref{sec:pl_prosody_modelling_ss} and \ref{sec:pl_prosody_modelling_ms}, the proposed system models the phone-level prosodies with GMM, which means the output of the prosody predictor is the parameters of GMMs. The other configurations of PLP-GMM are the same as PLP-SG.

\subsection{Prosody cloning}
We use LibriTTS dataset in our experiments on prosody cloning. The data preparation and acoustic model configurations are exactly the same as described in Section \ref{sec:conf_data_prep} and \ref{sec:conf_acoustic}. We select two speakers, one male and one female speaker, as the target speakers. The speech of the test set is synthesized with the two target speakers respectively, where the prosodies are cloned from the ground-truth speech of source speakers.

We build two systems in our experiments for prosody cloning: 1) the proposed GMM-based model and 2) the fine-grained VAE model \cite{prosody_transfer,copycat}.
The architecture of the fine-grained VAE model is reviewed in Section \ref{sec:copycat}. Instead of ULP that uses a single latent variable for the whole utterance, fine-grained VAE extracts prosody representations for short speech segments in order to precisely clone the prosodies from the reference. The latent variable in the experiment is 128 dimensional, which is the same as the prosody embeddings in PLP-GMM for fair comparison. To alleviate the speaker identity interference, \cite{copycat} introduces an instance norm layer in its prosody extractor and a bottleneck layer that downsamples the prosody representations with a factor $\tau$. In our experiments, we set the $\tau$ to 10.

\section{Experimental results and analysis}
\label{sec:exp_result}

\subsection{The necessity of using phone-level prosodies}
\label{sec:pl_vs_ul}
 Firstly, we verify whether using the extracted ground-truth phone-level prosodies $\mathbf{e}$ is better than using utterance-level prosodies in reconstruction. 
 Here, we reconstruct the test set with PLP-GMM and ULP, which is guided by the extracted ground-truth phone-level prosodies and by the utterance-level prosodies sampled from the posterior respectively.

Mel-cepstral distortion (MCD) \cite{mcd} is an objective measure of the distance between two sequences of mel-cepstral coefficients. We extract 25 dimensional mel-cepstral coefficients with 5ms frame shift \cite{mcd_25d} from both the synthetic speech and the corresponding ground-truth speech on the test set for computing the MCD. The results are demonstrated in Table \ref{tab:reconstruct}, where a lower MCD represents a better reconstruction performance. In both single-speaker(SS) and multi-speaker(MS) systems, using the extracted phone-level prosodies achieves lower MCD than the utterance-level baselines. This is natural because phone-level prosody representations contain much more information about how each phone is pronounced than an utterance-level representation. In addition, we also find that larger dimensionality of phone-level embeddings also provide slightly better results. Therefore, it is necessary to use phone-level prosodies in TTS systems for precise prosody controlling and we use 128 dimensional phone-level embeddings in the following experiments.

 \begin{table}[h]
\caption{Reconstruction performance (MCD) with utterance-level and phone-level prosody conditions on the test set.}
\label{tab:reconstruct}
\centering
\begin{tabular}{cc|cc}
\hline
  Granularity    &  Dim      &  SS      & MS       \\ \hline
Utterance-level  &  128     & 5.22            & 5.31     \\ 
Phone-level      &  32      & 3.56            & 3.68     \\ 
Phone-level      &  64      & 3.55            & 3.58     \\ 
Phone-level      &  128     & \textbf{3.38}   & \textbf{3.46} \\ \hline
\end{tabular}
\end{table}

\subsection{The number of Gaussian components for phone-level prosody modelling}
In this section, we try to figure out how many Gaussian components are needed to model the distribution of the extracted phone-level prosodies $\mathbf{e}$. We plot the log-likelihood curves on both the training set and the test set with several different numbers of Gaussian components in Figure \ref{fig:pitch_contour}. The observations in both the single-speaker and multi-speaker systems are similar. We can find that the log-likelihood gap between the training and test set in the single Gaussian model is larger than that in the Gaussian mixture models. Moreover, the log-likelihood curves of Gaussian mixture models are much higher than the single Gaussian model in both the training and test set. Therefore, we can conclude that the distribution of phone-level prosodies is multimodal and it should be modeled with Gaussian mixtures.

 \begin{table}[h]
\caption{The number of Gaussian components whose weights are greater than 0.1 and 0.01.}
\label{tab:activated_weights}
\centering
\begin{tabular}{c|cc}
\hline
             &  SS               &  MS     \\ \hline
$> 0.1$      & 3.44              & 4.78    \\ 
$> 0.01$     & 18.96             & 19.93   \\ \hline
\end{tabular}
\end{table}

Additionally, increasing the number of components also provides higher log-likelihood, because the more components enable the model to simulate more complicated distribution. However, when we double the number of the components from 10 to 20, the improvement is very limited. Therefore, we do not further increase the number of components, and use 20 components in the following GMM-based systems.

Then we synthesize the test set and analyze the predicted weights. We count the number of Gaussian components whose weights are greater than 0.1 and 0.01 for each phoneme. The average results are presented in Table \ref{tab:activated_weights}. We can find that the weights of nearly all 20 components are greater than 0.01, but only about 4 to 5 components are the majority ones whose weights are greater than 0.1.

 \begin{table}[h]
\caption{Objective evaluations of prosody diversity with MCD between synthetic speech with different sampled prosodies.}
\label{tab:prosody_diversity_mcd}
\centering
\begin{tabular}{c|c|c|c}
\hline
         & ULP   & PLP-SG & PLP-GMM \\ \hline
SS       & 2.51  &  4.15  & \textbf{4.47} \\ 
MS       & 2.65  &  4.28  & \textbf{4.59} \\ \hline

\end{tabular}
\end{table}

 \begin{table}[h]
\caption{Subjective evaluations of prosody diversity with AB preference listening test.}
\label{tab:prosody_diversity_ab}
\centering
\begin{tabular}{c|c|c|c}
\hline
                      & ULP    & PLP-SG   & PLP-GMM \\ \hline
\multirow{2}{*}{SS}   & 6.7\%  &  -       & 93.3\% \\ \cline{2-4}
                      & -      &  13.3\%  & 86.7\% \\ \hline
\multirow{2}{*}{MS}   & 8.0\%  &  -       & 92.0\% \\ \cline{2-4}
                      & -      &  14.0\%  & 86.0\% \\ \hline

\end{tabular}
\end{table}

\subsection{Prosody diversity}

According to the investigation above, we train the proposed system PLP-GMM with 20 Gaussian components. In the inference stage, the Gaussian mixture distributions of phone-level prosodies are predicted, from which the phone-level prosodies $\hat{\mathbf{e}}$ are sampled. 
The synthesis is then guided by the sampled prosodies $\hat{\mathbf{e}}$. \footnote{Audio examples are available here \texttt{\url{https://cpdu.github.io/gmm_prosody_modelling_examples}}.}

\begin{figure}[t]
\centering
\subfigure[Stressed rising tone]{
\includegraphics[width=\linewidth,height=2cm]{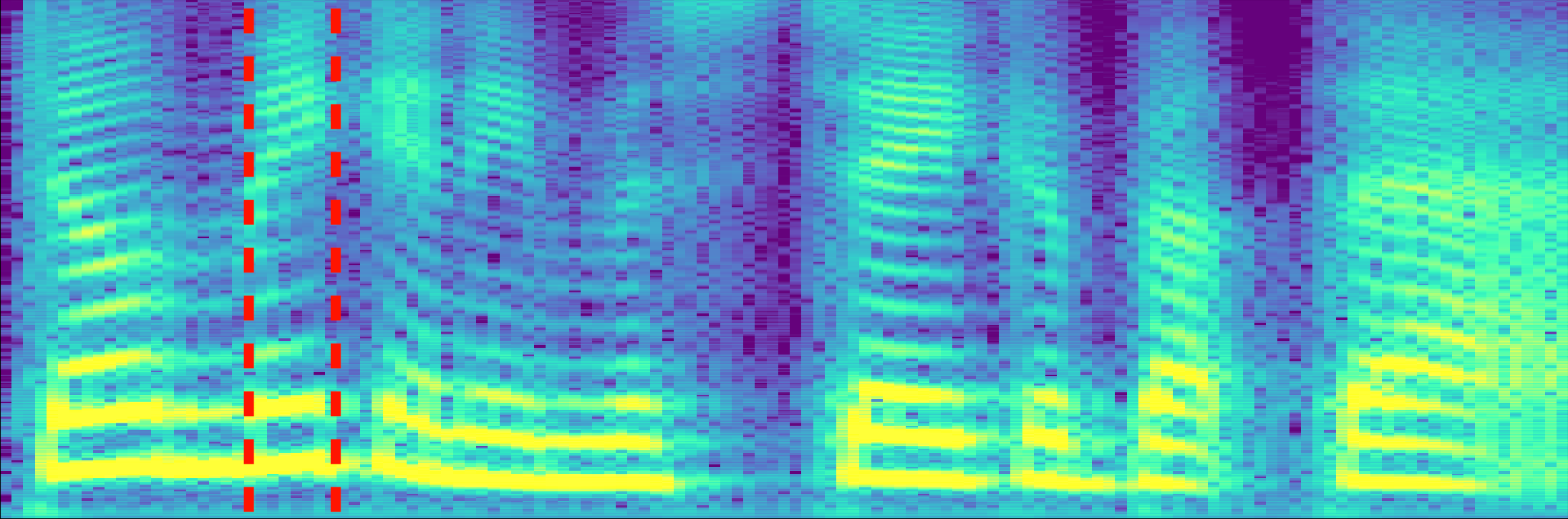}
}
\subfigure[Stressed falling tone]{
\includegraphics[width=\linewidth,height=2cm]{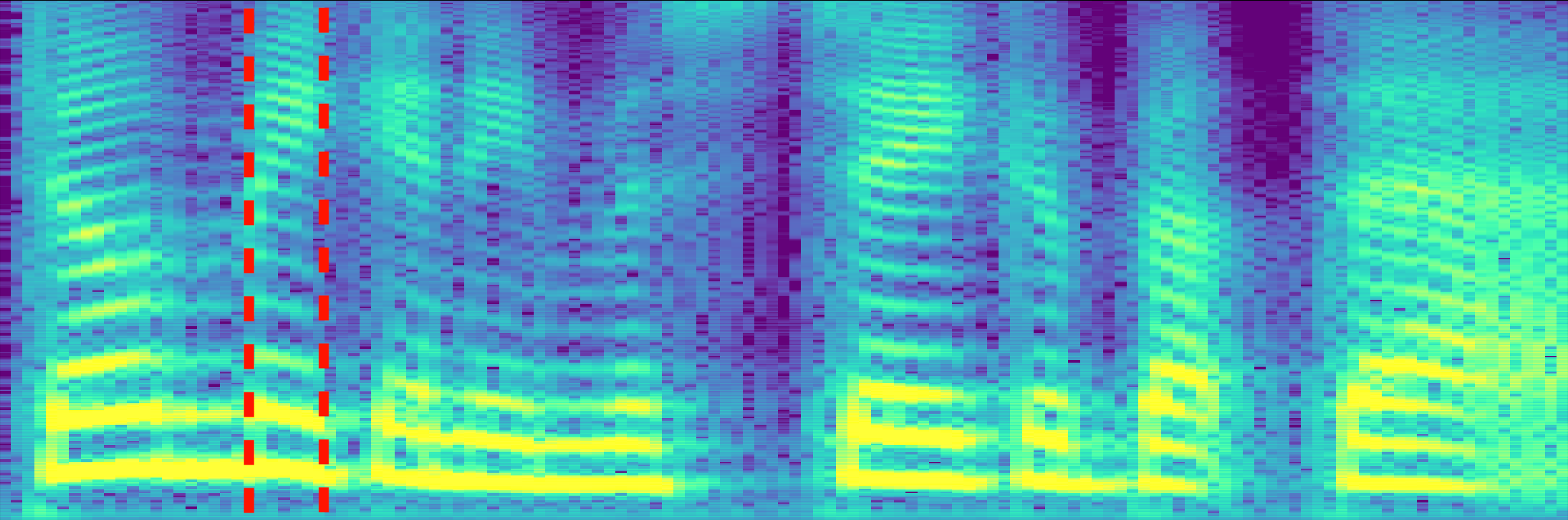}
}
\subfigure[Soft even voice]{
\includegraphics[width=\linewidth,height=2cm]{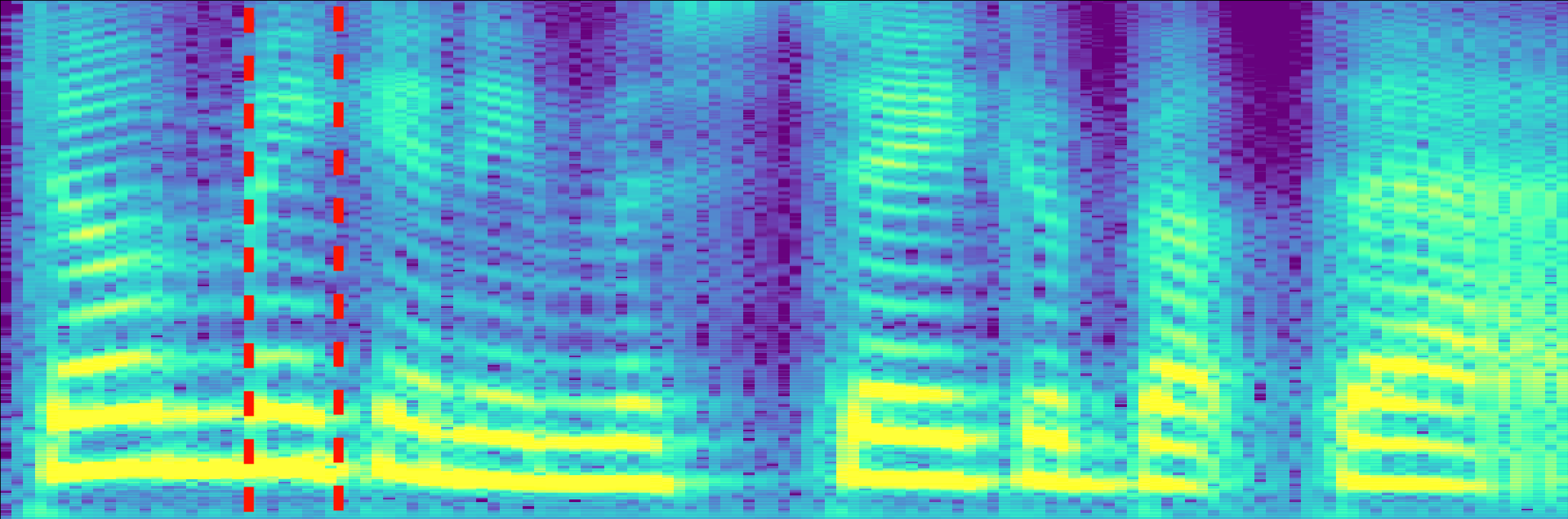}
}
\caption{The controllability of Gaussian index for phone \texttt{EY1} whose segments are between the red lines. The transcript is ``Don't make any mistake about that."}
\label{fig:idx_control}
\end{figure}

 We compare PLP-GMM with two baseline systems ULP and PLP-SG in terms of prosody diversity.
We synthesize each utterance in the test set 3 times with various sampled prosodies and calculate the average MCD between the 3 folds of synthetic speech. The results are presented in Table \ref{tab:prosody_diversity_mcd}, in which higher MCD means better prosody diversity. We also perform AB preference tests where two groups of synthetic speech from two different TTS systems are presented and 15 listeners need to select the better one in terms of prosody diversity. 10 groups of test cases are randomly selected from the test set for each listener.
Table \ref{tab:prosody_diversity_ab} show the results. We can find that PLP-GMM provides better prosody diversity in the synthetic speech than both ULP and PLP-SG.
This can be easily explained by the fact that a sequence of phone-level embeddings depicts the prosody more precisely than an utterance-level embedding and the fact that Gaussian mixtures can better model the phone-level prosody embeddings than a single Gaussian.

\subsection{Naturalness}

We also evaluate the naturalness of the above systems with a MUSHRA test, in which the 15 listeners are randomly presented with 10 groups of cases from the test set and are asked to rate each utterance on a scale of 0 to 100. The speech converted back from the ground-truth mel-spectrogram with the vocoder is also rated in the test, which is denoted as GT. 
 \begin{table}[h]
\caption{MUSHRA test in terms of naturalness with 95\% confidence interval.}
\label{tab:naturalness}
\centering
\begin{tabular}{c|c|c|c|c}
\hline
                     & PLP-GMM          & PLP-SG           & ULP             & GT            \\ \hline
SS   & $79.1\pm 1.6$  & $63.1\pm 1.8$  & $79.9\pm 1.3$ & $84.0\pm 0.8$ \\  \hline
MS   & $75.3\pm 0.9$  & $69.3\pm 1.6$  & $72.0\pm 1.3$ & $83.3\pm 1.0$ \\  \hline
\end{tabular}
\end{table}
The results are reported in Table \ref{tab:naturalness}. It can be observed that PLP-GMM synthesizes speech with better naturalness compared with PLP-SG because of the better phone-level prosody modelling. The MUSHRA score of PLP-GMM is 25.3\% and 8.6\% higher than PLP-SG on single-speaker and multi-speaker task respectively with p-value $< 0.05$. We can also find that ULP generates speech with similar naturalness as PLP-GMM, which can also be easily explained. ULP models the prosody only on utterance-level and no phone-level prosody is explicitly considered. The global prosody representation provide limited information on phone-level prosodies and is too coarse to precisely control the synthesis. In the sense of phone-level prosodies, the model uses averaged prosodies, whose naturalness is comparable to the sampled phone-level prosodies in PLP-GMM. However, both the naturalness of PLP-GMM and ULP are slightly lower than GT.

\begin{figure}[t]
\centering
\subfigure[Randomly sampled prosody]{
\includegraphics[width=0.8\linewidth]{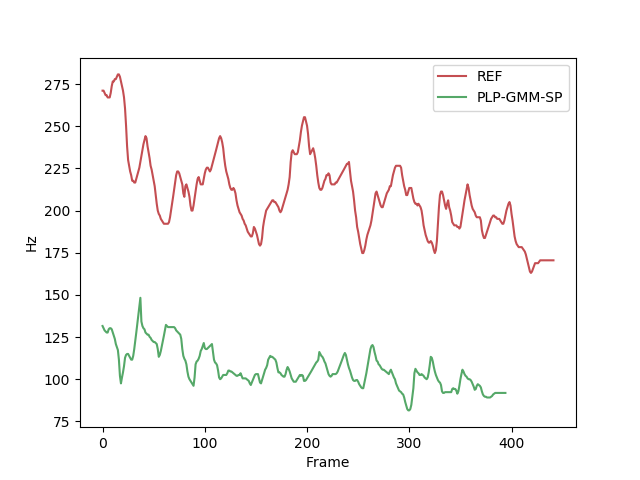}
\label{fig:randomly_sampled_prosody}
}
\subfigure[Cloned prosody]{
\includegraphics[width=0.8\linewidth]{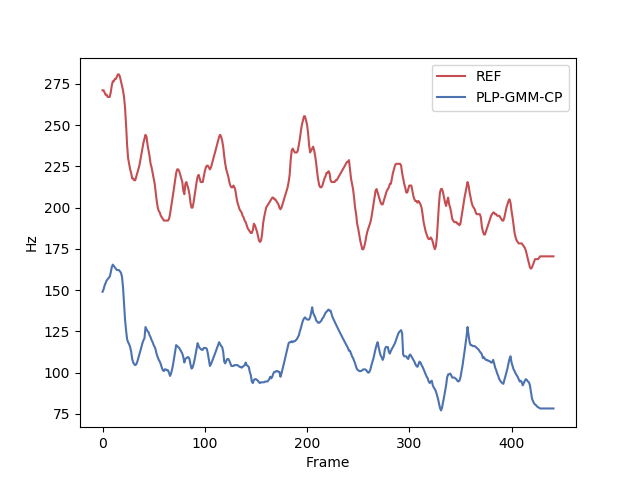}
\label{fig:cloned_prosody}
}
\caption{An example of pitch contours of the reference speech (REF), the synthetic speech with cloned prosody (PLP-GMM-CP) and the synthetic speech with randomly sampled prosody (PLP-GMM-SP). Here the reference speaker is female and the target speaker is male.}
\label{fig:pitch}
\end{figure}

\subsection{The controllability of Gaussian component index}

In GMM-based prosody cloning, we precisely control the synthesized phone-level prosodies with specified Gaussian indices. In this section, we explore whether Gaussian index could control the prosody. 
Figure \ref{fig:idx_control} illustrates 3 synthetic mel-spectrograms of the male target speaker, where only the prosody of phone \texttt{EY1} is sampled from 3 different Gaussian components while the prosodies of other phones are cloned from the specified components. In the figure, the segments of the phone \texttt{EY1} are between the red lines, where three different prosodies can be observed. Furthermore, the prosodies of other phones are almost the same, because they are all cloned from the reference speech. These results clearly show that the phone-level prosodies can be controlled by the Gaussian component indices.

\subsection{Pitch contour in GMM-based prosody cloning}
\label{sec:pitch_contour}
To visualize the GMM-based prosody cloning, we plot the pitch contour in this section, which is one of the most conspicuous factors of prosody. We extract pitch features with Kaldi \cite{kaldi} command \texttt{compute-kaldi-pitch-feats} for the reference speech (REF), the synthetic speech with cloned prosody (PLP-GMM-CP) and the synthetic speech with randomly sampled prosody (PLP-GMM-SP). Figure \ref{fig:pitch} illustrates an example where the reference speaker is female and the target speaker is male. We can observe that although the average pitch value of the female reference speaker is much higher than the target male speaker, the shape of the pitch contour is similar between PLP-GMM-CP and REF. The average pitch value of PLP-GMM-CP and PLP-GMM-SP is similar, because they are both the synthetic speech of the target male speaker, while the shapes of the pitch contour are quite different, because the prosody of PLP-GMM-SP is randomly sampled and is different from the reference.

\begin{figure}[t]
    \includegraphics[width=\linewidth]{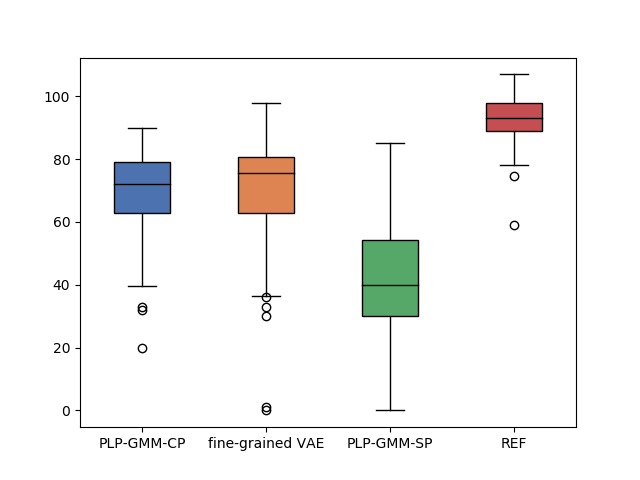}
    \caption{Boxplot of MUSHRA scores of prosody similarity.}
    \label{fig:prosody_similarity}
\end{figure}

 \begin{table}[t]
\caption{Prosody similarity evaluation with MUSHRA listening test with 95\% confidence interval and pitch correlation.}
\label{tab:prosody_similarity}
\centering
\begin{tabular}{c|c|c}
\hline
                     & MUSHRA score   & Pitch correlation \\ \hline
PLP-GMM-CP           & $69.2\pm 2.1$  & 0.58  \\ 
fine-grained VAE     & $70.8\pm 2.5$  & 0.51  \\ 
PLP-GMM-SP           & $42.2\pm 2.8$  & 0.28  \\
REF                  & $92.9\pm 1.1$  & 1     \\ \hline
\end{tabular}
\end{table}

\subsection{Evaluation of prosody cloning}

There are two goals in the prosody cloning task. On one hand, we need to clone the prosody from the reference speech of source speaker. On the other hand, we need to synthesize the speech with the specified target speaker without interference from the source speaker identity. Therefore, we evaluate the prosody similarity with the reference speech and the speaker similarity with the target speaker in this section.

In order to evaluate the prosody similarity between the synthetic speech and the reference speech, we consider using a MUSHRA test in which 15 listeners are randomly presented with 10 groups of cases and are asked to rate the prosody similarity on a scale of 0 to 100. In addition to PLP-GMM-CP and fine-grained VAE that applies prosody cloning, we also present PLP-GMM-SP and the reference speech in the test\footnote{Audio examples are available here \texttt{\url{https://cpdu.github.io/gmm_prosody_cloning_examples}}.}. The test results are reported in Figure \ref{fig:prosody_similarity} and Table \ref{tab:prosody_similarity}. As is expected, both PLP-GMM-CP and fine-grained VAE significantly improves the mean score by 64.0\% and 67.9\% respectively over PLP-GMM-SP that randomly samples the prosody, which demonstrate the effectiveness of prosody cloning. However, the reference still performs the best over both the prosody cloning methods. In addition to the subjective listening test, we also evaluate the pitch correlation between the synthetic speech and the reference. As is mentioned in Section \ref{sec:pitch_contour}, the pitch contour of the cloned prosody looks similar to that of the reference, so they should be positively correlated. We compute the average pitch correlation coefficients over all utterances in the test set and present the results in Table \ref{tab:prosody_similarity}. We can see the pitch of PLP-GMM-SP is not highly correlated to the reference, while PLP-GMM-CP and fine-grained VAE are positively correlated. Furthermore, the performance of PLP-GMM-CP and fine-grained VAE is comparable.

\begin{figure}[t]
    \includegraphics[width=\linewidth]{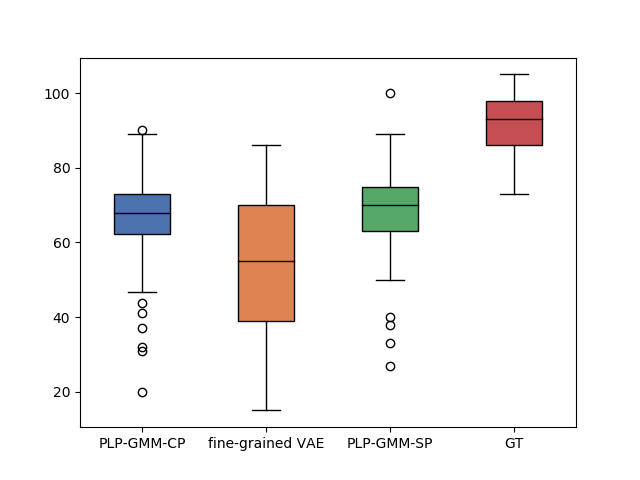}
    \caption{Boxplot of MUSHRA scores of speaker similarity.}
    \label{fig:speaker_similarity}
\end{figure}

 \begin{table}[t]
\caption{Speaker similarity evaluation with MUSHRA listening test with 95\% confidence interval and x-vector PLDA scoring.}
\label{tab:speaker_similarity}
\centering
\begin{tabular}{c|c|c}
\hline
                    & MUSHRA score       &  x-vector PLDA score \\ \hline
PLP-GMM-CP          & $66.1\pm 1.8$    &   -24.81 \\ 
fine-grained VAE    & $54.3\pm 2.9$    &   -30.13  \\ 
PLP-GMM-SP          & $68.2\pm 1.7$    &   -24.83  \\
GT                  & $91.9\pm 1.2$    &    21.52  \\ \hline
\end{tabular}
\end{table}


We also evaluate the speaker similarity in the MUSHRA test and illustrate the results in Figure \ref{fig:speaker_similarity} and Table \ref{tab:speaker_similarity}. It can be observed that the score of PLP-GMM-CP and PLP-GMM-SP is comparable. PLP-GMM-CP outperforms the fine-grained VAE with statistical significance ($p<0.05$) and improves the mean score by 21.7\%. We also evaluate the speaker similarity with x-vector PLDA scoring \cite{xvector}, where we compute log-likelihood ratios for the trials with a pretrained speaker verification model\footnote{\texttt{\url{http://kaldi-asr.org/models/m8}}.}. For each TTS system, the trial contains all its synthetic speech, including 2 folds of the test set with 2 target speakers respectively, and obtains speaker similarity scores between the synthetic speech and the enrollments of the corresponding target speakers. Here, the enrollments contain all the recordings of the target speakers in the training set. We average over the scores of all the utterances in the trials and report the results in Table \ref{tab:speaker_similarity}, which shows a similar conclusion with the subjective evaluation that PLP-GMM-CP and PLP-GMM-SP are comparable and they are both  better than the fine-grained VAE. This is as expected because the prosody representations in fine-grained VAE are prone to speaker identity interference.

\section{Conclusion}
\label{sec:conclusion}

In this paper, we propose a novel approach that utilizes a GMM-based mixture density network for phone-level prosody modelling in end-to-end speech synthesis. Then we further extend the GMM-based model for multi-speaker TTS with nonlinear transformations of Gaussian means and variances. Finally, we apply the GMM-based prosody modelling to prosody cloning task. Our experiments on LJSpeech and LibriTTS dataset show that the proposed GMM-based method not only achieves significantly better diversity than using a single Gaussian in both single-speaker and multi-speaker TTS, but also provides better naturalness. The prosody cloning experiments demonstrate that the prosody similarity of the proposed GMM-based method is comparable to recent proposed fine-grained VAE while the target speaker similarity is better.


\bibliographystyle{IEEEtran}
\bibliography{mybib}

\end{document}